# Coordinated Battery Electric Vehicle Charging Scheduling across Multiple Charging Stations

Saman Mehrnia[a*], Hui Song[a], Nameer Al Khafaf [a], Mahdi Jalili[a], Lasantha Meegahapola [a], Brendan McGrath[a]

[a] *School of Engineering, RMIT University, Melbourne, VIC, Australia*

**ABSTRACT**

The uptake of battery electric vehicles (BEVs) is increasing to reduce greenhouse gas emissions in the transport sector. The rapid adoption of BEVs depends significantly on the coordinated charging/discharging infrastructure. Without it, uncontrolled and erratic charging patterns could lead to increased power losses and voltage fluctuations beyond acceptable thresholds. BEV charge scheduling presents a multi-objective optimization (MOO) challenge, demanding a balance between minimizing network impact and maximizing the benefits for electric vehicle charging station (EVCS) operators and BEV owners. In this paper, we develop an MOO framework incorporating a carbon emission program and a dynamic economic dispatch problem, allowing BEV users to respond by charging and discharging through grid-to-vehicle (G2V) and vehicle-to-grid (V2G) technologies according to the optimal electricity price and compensation. Furthermore, we integrate dynamic economic dispatch with time-of-use tariffs to obtain optimal market electricity prices and reduce total costs over 24 hours. Our experimental results on a sample network show that the proposed scheduling increases participation in V2G services by over 10%, increases EVCS benefits by over 20%, and reduces network losses. Furthermore, increased rates of charging/discharging, coupled with more significant carbon revenue benefits for BEV users and EVCS, contribute to better offsetting battery degradation costs.

## 1. INTRODUCTION

Energy consumption across all sectors contributes significantly to global greenhouse gas emissions [1]. Battery electric vehicle (BEV) technologies within the transportation sector play a pivotal role in transitioning towards achieving net-zero carbon emissions [2, 3]. In recent years, there has been a notable surge in the rapid adoption of BEV numbers across numerous countries [4, 5]. Nevertheless, the increasing integration of BEVs can lead to unprecedented spikes in electricity demand. This can pose challenges to the power supply and infrastructure distribution network if BEV charging needs to be appropriately coordinated [6]. Moreover, uncoordinated BEV charging can increase the power system operating losses and costs due to the need for additional generation capacity to meet the increased demand during peak hours [7]. Implementing a coordinated charging strategy, especially when integrated with renewable energy systems (RESs), offers a promising approach to mitigating the potential adverse effects of unmanaged BEV charging and discharging activities [8]. This strategy not only ensures more efficient and sustainable use of RESs [9, 10] but also unlocks the potential of BEVs to support the grid, enhances the benefits beyond individual transportation, and transforms BEV owners into active participants in the grid [11, 12].

BEVs' grid-to-vehicle (G2V) and vehicle-to-grid (V2G) capabilities can enable coordinated charging/discharging [3]. These functionalities allow BEVs to draw power from the grid or supply power back to the grid when the BEV is idle. This occurs during high grid power demand periods or when the grid requires additional network support [12]. By leveraging these capabilities and optimizing scheduling and power management in EV charging stations (EVCSs), stakeholders, including BEV owners, distribution network service

---

[*] Corresponding author e-mail address: Mehrnia.Saman1992@gmail.com (Saman Mehrnia)

## NOMENCLATURE

### Abbreviations

| | | | |
|---|---|---|---|
| BDC | Battery degradation cost | MINLP | Mixed integer non-linear problem |
| BEV | Battery electric vehicle | MOO | Multi-objective function |
| DED | Dynamic economic dispatch | SOO | Single-objective function |
| DG | Distributed generation | NLP | Non-linear problem |
| DNSPs | Distribution network service providers | NSGA-II | Non-dominated Sorting Genetic Algorithm II |
| DR | Demand response | OEP | Optimal electricity price |
| DRM | Demand response management | P2V/PC | Peak to Valley / Peak Compensation |
| ERQ | Emission reduction quote | PEM | Price elasticity matrix |
| EVCS | Electric vehicle charging station | RES | Renewable energy sources |
| G2V | Grid-to-vehicle | SOC | State of charging |
| LF | Load factor | TOU | Time of use |
| SQPA | Sequential quadratic programming algorithm | V2G | Vehicle-to-grid |

### Notations

| | | | |
|---|---|---|---|
| $\alpha$ | Weight value | $R_{i,t}^{ERQ,EV}, R_{i,t}^{ERQ,CS}$ | The contributions of each BEV user and EVCS owner in the emission reduction quote |
| $C^b$ | The battery degradation cost | $S_i$ | The daily driving distance for the $i$th BEV |
| $c_i^b$ | BEV cost per kWh | $SOC_i^{ini}$ | Initial SOC for the ith BEV |
| $c^L$ | The labor cost of replacing a battery | $SOC_{i,t}^{min}, SOC_{i,t}^{max}$ | Lower and upper charging/discharging limits of $i$th BEV in each interval $t$. |
| $C^{TOU}$ | The cost associated with operationalizing the TOU program | $t_{out}^i$ | The departure time of the ith BEV |
| $c, d$ | Index of charging and discharging | $t_{back}^i$ | Return time of the ith BEV |
| $d_0, d$ | The demand before and after implementing the TOU program | $V^{min}, V^{max}$ | The minimum and maximum values of the voltage |
| $E_i^b$ | Battery Capacity for the $i$th BEV | $V_{j,t}$ | The voltage at each node $j$ in the network at each time interval $t$ |
| $E_i'$ | The energy required to travel one mile for the $i$th BEV | $N$ | Total number of BEVs |
| $E_{i,t}^{ERQ,EV}, E_{i,t}^{ERQ,CS}$ | The income generated from emission reduction credits associated with BEVs and EVCSs | $X_{i,t}^c, X_{i,t}^d$ | Charging or discharging status |
| $L$ | Battery life at a particular DOD | $(\sigma, \mu)$ | Mean and variance values for probability functions that characterize the behaviors of BEVs |
| $M$ | The number of network nodes $j \in \{1,…, M\}$, | $\pi^{loss}$ | The cost per kW of power losses |
| $P_{k,t}$ | Injected power from installed gas DGs across the network at a time slot $t$ | $\lambda_0, \lambda$ | The electricity price before and after implementing the TOU program |
| $P_t^{load}$ | The network load at a time slot $t$ | $\lambda_t^c, \lambda_t^d, \lambda_t^E$ | The charging, discharging price, and emission price |
| $P_t^{gid}$ | The injected power from the upstream to the downstream network at a time slot $t$ | $\eta_t^c, \eta_t^d$ | The battery efficiency rate of the charging and discharging |
| $P_t^{loss}$ | Power losses at each time interval $t$ | $\gamma^A, \gamma^B$, and $\gamma^C$ | Denote the percentage of BEVs in groups A, B, and C based on trip purposes |
| $P_{i,t}^c, P_{i,t}^d$ | Charging or discharging power for the ith BEV at the time slot t | $\rho$ | The electricity price adjustment factor |
| $P^{A,b}, P^{B,b}$, and $P^{C,b}$ | Pre-defined load profile of each group A, B, and C at the desired b-th node | $\rho_t^{min}, \rho_t^{max}$ | Minimum and maximum value for the adjustment factor |
| $P_t^{gid}$ | The injected power from the upstream to the downstream network at a time slot $t$ | | |

providers (DNSPs) and EVCS owners can benefit from reduced energy and operational costs, environmental advantages, improved voltage and frequency regulation, and peak power reduction. Optimization methods are essential for effectively addressing these scenarios [13-15]. In the context of BEV charging and discharging, optimal use refers to the strategic timing and management of charging/discharging activities to maximize economic, environmental, and operational benefits. This involves optimizing charging schedules to reduce costs, minimize peak demand, and enhance grid stability. Charge dynamics describe the variations in BEV charging behavior over time, influenced by factors such as electricity pricing, grid demand, and the availability of renewable energy sources. Understanding charge dynamics is crucial for developing effective strategies that optimize charging activities while mitigating their impact on the power grid.

Extensive research efforts have been dedicated to developing coordinated BEV charging strategies and analyzing these approaches' economic and operational trade-offs. For instance, a comprehensive review of BEV control structures in EVCS, power system objectives, and optimization methods for managing BEV charging and discharging activities is presented in [16, 17]. Furthermore, researchers in [18] implemented an economic dispatch model to reduce the cost of BEV users in microgrid systems. The proposed methods were validated through various case studies, showing that they effectively reduce costs, maximize energy savings, and encourage greater participation of BEV users in V2G services. Similarly, some works focused on minimizing BEV owners' charging costs while reducing demand from the upstream network and peak power consumption as part of an energy management system using linear and mixed integer programming [19-21]. A rules-based method on the state of charging (SOC) minimizes exchange power and improves local power stability [22]. Moreover, [22] presents a multi-objective scheduling model that effectively reduces EV charging costs and grid load variance through an improved hybrid algorithm. However, it fails to consider dynamic tariff optimization and lacks incentive policies for EV owners, limiting its adaptability and potential impact in real-world V2G applications.

While the studies mentioned above primarily focus on coordinated charging strategies using single-objective optimization (SOO) or combining multiple objectives into a single function, they often must address various competing objectives simultaneously. As a result, these approaches may overlook the complex trade-offs between maximizing EVCS profit, minimizing charging costs, reducing peak demand, enhancing grid stability, and reducing power losses. Therefore, in this context, each stakeholder, including the grid, BEV owners, and charging station operators, introduces distinct variables and conditions to the optimization problem [23]. The BEV coordination model's multi-objective function (MOO) has been implemented in [24] to minimize the average time, cost, and discrepancy between the expected and final SOC. The authors in [25] established a techno-economic-environmental optimization of BEVs for real-time electricity prices. The results show significant benefits with reduced costs, emissions, and improved grid utilization. Furthermore, a bi-level planning model is introduced and solved as a mixed integer non-linear problem (MINLP) by considering BEV users' autonomy and charging costs to optimize global economic costs while enhancing service satisfaction, incorporating different factors [26]. Similarly, researchers [27] proposed a MOO technique for BEV fleet charging using a non-linear problem (NLP), yielding an optimal profile that minimizes charging and battery aging costs while achieving a 20% reduction in electricity cost and a 48% reduction in battery aging. Despite the studies on MOO, implementing two-way charging remains a challenging problem, primarily because BEV owners are concerned about the additional costs associated with more frequent battery cycling in V2G services. In the study above, battery degradation cost (BDC) is seldom examined as a critical factor in multi-criteria approaches to decision-making for V2G and G2V services, which could otherwise provide significant economic and environmental benefits. Developing a holistic scheme that benefits BEV owners engaging in V2G services that consider BDC is substantial [28]. Additionally, incentive programs that encourage BEV owners to participate in V2G and G2V activities are rarely applied. There is also a noticeable gap in developing programs that ensure these incentives lead to tangible and beneficial outcomes for all stakeholders involved.

Specific investigations considered the BDC [28-33] to reduce the cost associated with BEV charging, while others overlooked the consequences of battery degradation [19, 29]. In contrast, [30] deployed an advanced battery management system to optimize the G2V/V2G activities, enhancing battery longevity. Another study [31] links battery degradation with distribution planning, suggesting that considering the depth of discharge impacts can improve asset management and battery lifespan. Similarly, [32] uses a MILP cost function to optimize the trade-off between demand charge reduction and energy arbitrage, revealing that deep and frequent cycling significantly shortens the BESS lifespan. However, these studies do not introduce incentive schemes to encourage BEV owners to participate in two-way charging programs. Some research focuses on dynamic electricity pricing as a mechanism to reduce operational costs and motivate BEV owners [33-35]. For instance, time of use (TOU) pricing is the most common structure to effectively reduce total energy costs by encouraging charging during off-peak and discharging during peak periods [36]. Optimal electricity pricing (OEP) can make V2G services economically viable, offsetting ownership costs and potentially generating additional income for BEV owners. Another study [37] applied a demand response (DR) program, reducing charging costs and improving distribution network load variance. Additionally, implementing a dynamic pricing mechanism based on peak load conditions, integrated with an adaptive decision-making process, can optimize scheduling by minimizing charging costs during off-peak hours [13]. Similarly, [38] proposed a two-level BEV charging scheduling method at EVCS, utilizing a dynamic pricing strategy to shift BEV loads from peak to off-peak hours, thereby reducing charging costs and alleviating strain on the grid. While some studies focus on TOU tariffs, few explore subscription plans offering additional benefits like free charging sessions or priority access to charging infrastructure [39, 40]. Nevertheless, these studies fail to provide a practical scheme to address BDC comprehensively or compensate for the extra cost of battery degradation.

The following research gaps have been identified based on the current literature:

- Integration of competing objectives: Existing studies often focus on single-objective or simplified multi-objective optimization for BEV charging strategies. However, they generally overlook the need to balance competing objectives, such as EVCS profit, BEV owner costs, and network power losses. This gap highlights the need for a more comprehensive MOO framework to balance these competing interests effectively.
- Battery degradation cost inclusion: While some studies consider BDC in their models, many fail to comprehensively incorporate it into decision-making processes for V2G and G2V services. There is a noticeable gap in developing strategies that adequately address BDC, particularly in a way that provides tangible benefits to BEV owners engaging in V2G activities.
- Lack of effective incentive programs: Current research often neglects the development and implementation of incentive programs that encourage BEV owners to participate in V2G services. There is a significant gap in creating practical schemes that not only motivate participation but also ensure that these incentives lead to beneficial outcomes for all stakeholders involved.
- Dynamic pricing mechanisms: Although dynamic pricing mechanisms like TOU tariffs have been explored, limited research exists on alternative pricing structures, such as subscription plans offering supplementary benefits. This gap indicates the need for innovative pricing models to better align with BEV owners' interests and promote sustainable grid integration.
- Comprehensive BEV charging frameworks: Existing studies often lack a holistic approach to BEV charging strategies that integrate various factors, such as load profile variability, uncertainty in PV generation, and varying charging/discharging rates. This gap underscores the necessity of developing more robust frameworks that can adapt to the diverse needs of BEV owners while optimizing grid performance.

This paper addresses the identified research gaps by proposing a comprehensive framework that optimizes BEV charging and discharging activities while considering the complex interactions between various stakeholders and critical factors like battery degradation and load balance profile. The key contributions of this study are:

- Development of an MOO Framework: We introduce a MOO framework that simultaneously addresses three competing objectives: maximizing EVCS profit, minimizing network power losses, and reducing BEV owners' costs. This approach provides a balanced optimization that caters to all stakeholders' needs.

- Integration of diverse BEV load profiles: The framework incorporates load profiles from different categories of BEV owners, considering diverse driving patterns and varying charging/discharging rates. This integration ensures that the model reflects real-world scenarios and captures the uncertainty in load profiles and PV generation, thereby improving the reliability of the optimization outcomes.
- Implementing dynamic economic dispatch (DED) and demand response management (DRM): The study proposes integrating DED and DRM strategies to generate an optimal electricity pricing signal. This innovative approach encourages BEV owners to participate in V2G services and optimize their load profile patterns, resulting in cost reductions and maximized benefits.
- Introduction of a compensation mechanism on carbon credits: A novel compensation mechanism is proposed to address battery aging costs, which is often a concern for BEV owners. This mechanism converts carbon credits into monetary values, providing a tangible incentive for BEV and EVCS owners to participate in V2G services, thereby aligning economic benefits with environmental sustainability.
- An empirical comparison was conducted between the BEV charging/discharging scheduling method proposed in this paper and the method in [25], which employs a sequential quadratic programming (SQP) algorithm for multi-objective techno-economic-environmental optimization of BEVs. Additionally, a comparison was made with the dispatch method in [41, 42], which utilizes a water-filling algorithm to allocate BEV load before dispatch. While the grid load impact of the proposed method is similar to both alternative methods, our approach provides greater convenience for BEV users to participate in V2G services. It reduces concerns about frequent battery cycling in V2G and G2V operations.

Table 1 compares the coordinated BEV charging design model developed in this study and previous models found in the literature. It outlines the key contributions and improvements introduced by the proposed model. The table highlights how various innovative features have been incorporated, setting this model apart from earlier approaches and enhancing its effectiveness in addressing current BEV charging challenges.

Table 1: Comparative analysis of the proposed coordinated BEV charging design model and previous studies

| Model characteristics | [12] | [13] | [14] | [15] | [25] | [26] | [31] | [32] | [37] | [43] | [44] | Proposed |
|---|---|---|---|---|---|---|---|---|---|---|---|---|
| Multi-objective optimization problem | ✗ | ✗ | ✓ | ✗ | ✓ | ✓ | ✗ | ✗ | ✓ | ✗ | ✓ | ✓ |
| Optimal charging strategies for BEVs | ✗ | ✓ | ✓ | ✗ | ✗ | ✓ | ✗ | ✓ | ✗ | ✗ | ✓ | ✓ |
| Optimal G2V/V2G services | ✓ | ✗ | ✗ | ✓ | ✓ | ✗ | ✓ | ✗ | ✓ | ✓ | ✗ | ✓ |
| Cost of battery degradation | ✗ | ✗ | ✓ | ✗ | ✓ | ✓ | ✗ | ✓ | ✓ | ✓ | ✗ | ✓ |
| Uncertainty in BEV owner behavior and network elements | ✗ | ✗ | ✓ | ✗ | ✗ | ✓ | ✓ | ✓ | ✗ | ✗ | ✓ | ✓ |
| Environmental impact | ✗ | ✗ | ✗ | ✗ | ✓ | ✗ | ✗ | ✗ | ✗ | ✗ | ✓ | ✓ |
| Charging type (fast, regular, slow) | ✗ | ✗ | ✗ | ✗ | ✗ | ✗ | ✗ | ✗ | ✗ | ✗ | ✗ | ✓ |
| Profits for EVCS | ✗ | ✗ | ✗ | ✗ | ✗ | ✓ | ✗ | ✗ | ✗ | ✗ | ✓ | ✓ |
| Optimal electricity price | ✗ | ✓ | ✗ | ✗ | ✗ | ✗ | ✗ | ✗ | ✗ | ✗ | ✗ | ✓ |
| Power losses in the network | ✗ | ✗ | ✗ | ✗ | ✗ | ✗ | ✗ | ✗ | ✗ | ✗ | ✗ | ✓ |
| Participation-driven V2G incentive program | ✗ | ✗ | ✗ | ✗ | ✗ | ✗ | ✗ | ✗ | ✗ | ✗ | ✗ | ✓ |

The subsequent sections of this study are structured as follows: Section 2 provides the problem modeling. Section 3 introduces the proposed method and its MOO implementation. Section 4 shows the simulation results and analysis of the introduced scenarios. Finally, Section 5 presents the conclusions, summarizes the study's key findings, and outlines potential future work, including the limitations of the current research and areas for further exploration.

## 2. PROBLEM MODELING

Initially, we will present a comprehensive BEV fleet modeling procedure designed to allocate BEVs across the network nodes efficiently. Subsequently, we will meticulously delineate the roles of each participant within an MOO framework. This exploration will delve into the intricate dynamics of the network's impacts on the coordinated 24-hour charging and discharging schedule for BEVs while concurrently considering the emission reduction program as a pivotal compensation mechanism. Finally, we will present a cost reduction program that combines dynamic economic dispatch problems with demand response programs.

## 2.1 BEV fleet modeling

Before introducing the problem, we will present the modeling of BEV users' traffic behavior and then explain how to assign BEVs to each network node. BEV driving behavior can vary based on individual preferences, activities, needs, and circumstances and can be categorized into three groups. It is essential to include stochastic assessment to depict the behavior of BEVs concerning uncertain variables accurately. We employ the Monte Carlo Method to address the stochastic BEV owners' driving behavior. We adjust the simulation settings to capture actual driving patterns accurately. We identify four variables, departure time, return time, daily driven distance, and initial SOC, which are crucial for modeling the driving pattern. The Monte Carlo method generates the probability density function of the output variable through iterative trials with randomly generated input variables. It can handle non-linear systems without requiring complex linearization methods and serves as a benchmark for assessing the accuracy of other probabilistic computational techniques. The departure time ($t_{out}$), return time ($t_{back}$), daily driving distance ($s$), and $SOC_i^{ini}$ the initial SOC for individual BEV are generated according to the following probability distributions [18]:

$$
\begin{aligned}
f(t_{out}) &= \frac{1}{\sqrt{2\pi}(\sigma_1)} e^{\left(-\frac{(t_{out}-\mu_1)^2}{2\sigma_1^2}\right)}, \quad t_{out} \geq 0 \\
f(t_{back}) &= \frac{1}{\sqrt{2\pi}(\sigma_2)} e^{\left(-\frac{(t_{back}-\mu_2)^2}{2\sigma_2^2}\right)}, \quad t_{back} \geq 0 \\
f(s) &= \frac{1}{\sqrt{2\pi}(\sigma_3 s)} e^{\left(-\frac{(Ln(s)-\mu_3)^2}{2\sigma_3^2}\right)}, \quad s \geq 0 \\
SOC_i^{ini} &= \left\{\left(\frac{E_i^b - s_i \times E_i'}{E_i^b}\right)\right. \quad E^b, E' > 0
\end{aligned}
\tag{1}
$$

where subscripts ($\sigma_1$, $\sigma_2$, $\sigma_3$) and ($\mu_1$, $\mu_2$, $\mu_3$) represent the mean and variance values of three distinct probability functions that characterize the behaviors of BEVs. The subscript $i$ represents the BEV index vector $\forall i \in \{1,\ldots,N\}$, and $N$ represents the total number of BEVs. $SOC_i^{ini}$, $s_i$, $E_i^b$, and $E_i'$ represent the initial SOC, driving distance, battery capacity, and energy required to travel one mile for the $ith$ BEV. We set $\mu_1=7$, $\mu_2=17$, $\mu_3=3.2$, $\sigma_1=1$, $\sigma_2=2$, and $\sigma_3=0.88$.

The number of BEVs connected to each node depends on the load type. This study incorporates the load consumption patterns of all BEV owners, classified into three groups: A, B, and C.

- Group A: BEV drivers who use their cars sporadically for shorter distances, such as non-working parents and unemployed individuals engaging in quick shopping or personal activities.
- Group B: BEV drivers with specific driving behaviors and fixed commuting hours, such as office workers who commute to and from work during designated times.
- Group C: BEV drivers with unpredictable daily consumption patterns, such as tradespeople and delivery personnel covering long distances.

By categorizing independent BEV owners into these three groups, estimating the number of BEV users within each group is feasible, given the system's total number of BEV owners. Assigning BEV load per node involves evaluating the total load contributed by BEVs connected to each node within the grid. This assessment includes analyzing the charging demands, load profiles, and power requirements of the BEVs associated with a particular node. The impact of BEV activity on the network can be determined by assessing the BEV load per node, enabling informed decisions on grid management and load-balancing strategies. The estimation of BEV loads in each node is formulated as follows:

$$n^{A,b} = N \times \left[ \frac{\Upsilon^A \times P^{A,b}}{\sum_{s=2,5} P^A} \right], n^{B,b} = N \times \left[ \frac{\Upsilon^B \times P^{B,b}}{\sum_{s=4} P^B} \right], n^{C,b} = N \times \left[ \frac{\Upsilon^C \times P^{C,b}}{\sum_{S=1,3} P^C} \right] \quad (2)$$

where the subscripts $A$, $B$, $C$, and $b$ represent groups A, B, and C, and node index, s is the network segment. $\gamma^A$, $\gamma^B$, and $\gamma^C$ denote the percentage of BEVs in groups A, B, and C for trip purposes. $P^{A,b}$, $P^{B,b}$, and $P^{C,b}$ indicate the pre-defined load profile of the desired node based on each determined segment. $\sum^{n_b} P^A$, $\sum^{n_b} P^B$, and $\sum^{n_b} P^C$ denote the total pre-defined load profile of the nodes based on IEEE network data aimed at group A, B, and C trip purposes.

### 2.2 Coordinated BEV charging modeling

The two-way charging aims to minimize BEV owners' costs during their 24-hour charging and discharging activities within EVCSs. This objective encompasses four different components, each of which plays a crucial role in the overall cost structure:

- The electricity tariffs associated with peak, off-peak, and shoulder-hour rates determine the BEV charging expenses.
- The income from BEV discharge varies depending on the electricity prices during the V2G process.
- A carbon credit mechanism reduces costs for BEV owners during the discharge period and enhances financial prospects for both BEV and EVCS owners. It promotes BEV participation in V2G without concerns about increased charging/discharging costs.
- Wear and tear from repeated charging and discharging cycles causes BDCs. Accurately accounting for these costs is essential for assessing BEV ownership's long-term viability and cost-effectiveness.

By considering and optimizing these four components, the two-way coordinated BEV charging aims to create a balanced and economically sustainable power network for BEV users, defined as follows:

$$Min \quad C = \sum_{t_i^{back}}^{t_i^{out}} \left( X_{i,t}^c . \lambda_t^c . P_{i,t}^c . \eta_i^c - X_{i,t}^d . \lambda_t^d . P_{i,t}^{d,p} . \eta_i^d - R_{i,t}^{ERQ,EV} \right) \Delta t + C^b \quad (3)$$

here, the subscripts $c$ and $d$ represent charging and discharging, respectively. $t_{out}^i$ and $t_{back}^i$ indicate the returning departure time and departure time of the $i$th BEV, respectively. $X_{i,t}^c$, $X_{i,t}^d$, $p_{i,t}^c$ and $P_{i,t}^d$ denote charging or discharging status, trading power for each BEV at the time slot $t$. $\lambda_t^c$, $\lambda_t^d$, $\eta_i^c$ and $\eta_i^d$ indicate the charging and discharging price and the battery efficiency of the charging and discharging, respectively. $R_{i,t}^{ERQ.EV}$ indicates the contributions of each BEV user in the emission reduction quote, and $C^b$ is the BDC. Moreover, each BEV operates in a specific state, charging, discharging, or idle during each time interval at the EVCS. These states must be considered for each BEV and are modeled as follows:

$$|X_{i,t}^c| + |X_{i,t}^d| \leq 1 \quad , \begin{cases} X_{i,t}^c \in \{0,1\} \\ X_{i,t}^d \in \{-1,0\} \end{cases} \quad (4)$$

Additionally, in this study, to implement a compensation mechanism known as carbon credits earned by BEV and EVCS owners, the following assumptions are considered:

- The thermal gas power plant is the sole external power source, supplementing the electricity demand alongside PV generation.
- The efficacy of emission reduction of EVCSs equipped with PV canopies is significant compared to thermal units generating an equivalent amount of power as PV systems, known as emission reduction quotes (ERQ).
- The emission reduction assigned to BEV owners reduces pollution during the V2G service's discharging process. It reflects the pollution generated by thermal units producing equivalent power during this phase.

To obtain a compensation mechanism for BEV owners, we consider the following equation:

$$E_{i,t}^{ERQ,EV} = \sum_{i=1}^{N} \sum_{t=1}^{T} p_{i,t}^{d} \times E^{gas} \tag{5}$$

where $E_{i,t}^{EV}$, $E^{gas}$, $P_{i,t}^{d}$ are the emission reduction related to BEV, the emission factor per unit mileage of fuel vehicles, and the discharging power of *i*th BEV at time interval t, respectively. By obtaining this credit, they can sell it in the carbon credit market and turn it into revenue. We use Eq. (6) to transfer this credit to dollars:

$$R_{i,t}^{ERQ,EV} = \sum_{i=1}^{N} \sum_{t=1}^{T} E_{i,t}^{ERQ,EV} \times \lambda_{t}^{E} \tag{6}$$

where $R_{i,t}^{ERQ.EV}$, and $\lambda_{t}^{E}$ indicate the revenue of BEV users from the ERQ mechanism and emission price, respectively. The degradation cost of battery function is formulated as follows [41]:

$$C^{b} = \sum_{i=1}^{N} \left( c_{i}^{b} \times E_{i}^{b} + c^{L} \right) \times P_{i}^{d} / \left( L \times E_{i}^{b} \times DOD \right) \tag{7}$$

where $L$ is the battery life at a particular depth of discharge (DOD), $E_{i}^{b}$ and $c_{i}^{b}$ refer to the capacity of each BEV and the BEV cost per kWh, $c^{L}$ represents the labor cost, $P_{i}^{d}$ is the discharging power of the ith BEV, and DOD presents the depth of discharge. The charging state of BEVs at any control time interval should satisfy a minimum and maximum range. At each time slot, the charging state of the *i*th BEV depends on its previous SOC and the efficiency of charging/discharging, as defined as follows:

$$SOC_{t,i}^{\min} \leq SOC_{t,i} \leq SOC_{t,i}^{\max}, \quad 0 \leq SOC_{t,i} \leq E_{i}^{b} \tag{8}$$

$$SOC_{i,t} = SOC_{i,t-1} + \sum_{t_{i}^{back}}^{t_{i}^{out}} \left( X_{i,t}^{c} \times \frac{\eta_{i,t}^{c} \times p_{i,t}^{c}}{E_{i}^{b}} - X_{i,t}^{d} \times \frac{\eta_{i,t}^{d} \times p_{i,t}^{d}}{E_{i}^{b}} \right) \Delta t \tag{9}$$

where $SOC_{i,t}^{min}$, $SOC_{i,t}^{max}$ denote the lower and upper charging/discharging limits of *i*th BEV in each interval *t*. The second term of Eq. (9) represents the change in the BEV battery SOC at each time slot *t* during G2V and V2G services. To enhance battery longevity, it is better to maintain SOC between 20% and 90% via battery management system control [45], or it can be operated between 10% and 95% [46]. However, according to [47, 48], this study considers an SOC range between 5% and 95% as an input parameter, as defined in Eq. (8).

### 2.3 EVCS benefits

The first objective is maximizing the EVCS owners' benefits over a 24-hour by providing optimal and cost-effective energy for V2G and G2V services. It includes three distinct terms, each crucial to the overall benefit structure:

- The first term relates to the income earned from selling energy by providing charging services. This energy can be sourced externally from gas-distributed generations (DGs), the upstream network, or internally, like PV systems and the discharging energy of other BEVs.

- The second term involves the income generated from having a PV canopy system, which provides emission reduction credits from the PV system.
- The third term refers to the cost of purchasing energy from BEVs through V2G services. EVCS owners can buy surplus energy from BEVs by enabling bidirectional energy flow.

$$\text{Max} \quad f_1 = \sum_{i=1}^{N} \sum_{t=1}^{T} \left( (p_{i,t}^c \times \lambda_t^c) + R_{i,t}^{ERQ,CS} - (p_{i,t}^d \times \lambda_t^d) \right) \quad p_{i,t}^c, p_{i,t}^d \geq 0 \tag{10}$$

$$\sum_{i=1}^{N} P_{t,i}^c = P^{grid} + \sum_{k=1}^{N^{DG}} P_{k,t} + \sum_{i=1}^{N} P_{t,i}^d + \sum_{j=1}^{N^{CS}} P_j^{PV} \tag{11}$$

$$E_{i,t}^{ERQ,CS} = \sum_{j=1}^{N^{CS}} \sum_{t=1}^{T} p_{j,t}^{PV} \times E^{th} \tag{12}$$

here, $E_{i,t}^{ERQ,CS}$, $P_{j,t}^{PV}$, and $E^{th}$ are the emission reductions related to EVCS, the amount of power PV generates, and the emission factor per power generation unit. $P^{grid}$ refers to the injected power from the upstream to the downstream network, $P_{k,t}$ represents the injected power from installed gas DGs across the network. To convert this credit into monetary value, we consider the following equations:

$$R_{i,t}^{ERQ,CS} = \sum_{i=1}^{N^{CS}} \sum_{t=1}^{T} E_{i,t}^{ERQ,CS} \times \lambda_t^E \tag{13}$$

where $R_{i,t}^{ERQ,EV}$ and $\lambda_t^E$ indicate the revenue of the charging station from the emission reduction mechanism and emission price, respectively.

### 2.4 Grid power losses

Minimizing grid power losses involves strategically managing BEVs' charging/discharging activities within the network. This includes scheduling the charging of BEVs during off/shoulder peak hours when the network capacity is underused and discharging energy back to the network during peak hours to support the load. Additionally, incorporating RESs, such as PV systems, can reduce the dependency on conventional power sources, thereby reducing transmission losses. The optimal operation of EVCSs also plays a vital role in minimizing power losses, ensuring that the energy flow is more direct and requires less transmission distance, reducing power losses. Minimizing grid power losses not only enhances the efficiency of the energy distribution but also contributes to the sustainability and cost-effectiveness of the power system. This results in a more resilient grid that can better accommodate the increasing penetration of BEVs and RESs. We assume that operation and maintenance (OM) costs are zero. The grid power loss is defined as:

$$\text{Min} \quad f_2 = \sum_{t=1}^{T} \left( \pi^{loss} \times P_t^{loss} \right) \tag{14}$$

$$V^{min} \leq V_{j,t} \leq V^{max} \quad \forall j \in \{1,...,M\}, \quad \forall t \in \{1,...,T\} \tag{15}$$

where $\pi^{loss}$ denotes the cost per kW of power losses. $P_t^{loss}$ power losses at each time interval $t$, which depend on the network topology, can be described as shown in [49]. $V^{min}$ and $V^{max}$ denote the minimum and maximum values of voltage. We consider the IEC 60038-2009 standard requires the node voltage to be -6% - +10% of the nominal system voltage, $V_{j,t}$ refers to the voltage at each node $j$ in the network at each time interval $t$, $M$ denotes the number of network nodes $j \in \{1,..., M\}$, $t$ refers to the time interval.

## 2.5 DED and DRM modeling

We combine demand response and dynamic economic dispatch programs. The DR approach reduces costs for prosumers by focusing on the demand side and considering the elasticity of charging/discharging demand. In contrast, using a quadratic cost model for generation units [50, 51], the DED problem minimizes overall generation costs, allowing for accurate cost representation. Incorporating this model into coordinated charging benefits BEV users and EVCS owners, achieving cost-reduction objectives. The cost-benefit analysis considers the costs of implementing and operating the TOU program versus its benefits, such as reduced peak demand, improved load balancing, enhanced network reliability, and potential revenue from DR program participation. The cost associated with operationalizing the TOU is evaluated as follows:

$$C^{TOU}(t) = \lambda(t) \times d(t) - \lambda_0(t) \times d_0(t) \tag{16}$$

here $\lambda_0(t)$, $d_0(t)$, $\lambda(t)$, and $d(t)$ represent the electricity price and demand before and after implementing the TOU program. To derive $\lambda$ during three different periods, the parameter $\rho$ is utilized:

$$\lambda = \begin{bmatrix} \lambda_{peak} \\ \lambda_{Valley} \\ \lambda_{off-peak} \end{bmatrix} = \begin{bmatrix} \lambda'_{peak} + \rho \\ \lambda'_{Valley} - \rho \\ \lambda'_{off-peak} \end{bmatrix} \tag{17}$$

where $\lambda'_{peak}$, $\lambda'_{Valley}$, and $\lambda'_{off-peak}$ refer to initial electricity prices during different periods, and $\lambda$ denotes the updated value of electricity prices. $\rho$ refers to the peak-period price adjustment factor, which adjusts electricity prices during peak and off-peak times. A line search method is applied to obtain the $\rho$ optimal value [52]. The line search method is a numerical optimization technique that iteratively finds the optimal step size along a chosen direction to minimize or maximize an objective function. By evaluating the function along this direction, it adjusts the step size to ensure the function value decreases or increases as desired. As $\rho$ increases, the $\lambda_{peak}$ rises while the $\lambda_{Valley}$ decreases, continuing until the optimal $\rho$ value is found. This process minimizes the $C^{TOU}$ (Eq. (16)) and establishes the optimal $\rho$ for the TOU scheme (Eq.(17)). The goal is to incentivize customers to shift their consumption from peak periods to valley or off-peak times, thereby balancing demand. In this study, we considered a range for $\rho$ from 0 to 20, with increments of 0.25. We implemented minimum and maximum limits for the adjustment factor to ensure that electricity prices remain within a reasonable range, as shown in Eq.(18).

$$\rho_t^{min} \leq \rho_t \leq \rho_t^{max} \tag{18}$$

To reduce BEV costs, users should optimize their charging behavior by charging during off-peak times when electricity prices are lower and discharging during peak times when prices are higher, effectively responding to demand response signals. Therefore, to impact the load patterns of BEV owners, elasticity is defined as the responsiveness of demand to price, is described as follows:

$$E(t,t') = \frac{\lambda_0(t')}{d_0(t)} \times \frac{\partial d(t)}{\partial \lambda(t')} \begin{cases} E(t,t') \leq 0 & if \quad t = t' \\ E(t,t') \geq 0 & if \quad t \neq t' \end{cases} \tag{19}$$

The net BEV user's profit function as Eq. (20) is introduced to achieve the modified consumption pattern. By taking the derivative of the desired function, we have:

$$NP(t) = P(d(t)) - d(t)\lambda(t) \tag{20}$$

$$\frac{\partial NP(t)}{\partial d(t)} = 0 \rightarrow d(t) = d_0(t) \times \left\{ 1 + \sum_{t=1}^{24} E(t,t') \frac{\lambda(t) - \lambda_0(t)}{\lambda_0(t)} \right\} \tag{21}$$

where *NP(t)* and *P* denote the net profit and the prosumer profit obtained by consuming power after implementing the TOU program, respectively.

The network balance equation is expressed as:

$$P_t^{grid} + \sum_{k=1}^{N^{DG}} P_{k,t} + P_t^{PV} = \sum_{i=1}^{N} P_{i,t}^c - \sum_{i=1}^{N} P_{i,t}^d + P_t^{load} + P_t^{loss} \qquad P_{i,t}^c, P_{i,t}^d \geq 0 \qquad (22)$$

here $P_t^{gid}$, $P_{k,t}$, $P_t^{PV}$ and $P_t^{load}$ represent the following variables at a time slot *t*: the injected power from upstream to the downstream network injected power from installed gas DGs across the network, injected power from PV systems, and the network load, respectively.

We apply the price elasticity matrix (PEM) to assess flexible demand potential in DR by measuring customer sensitivity to price changes. Defined in Eq. (23), the PEM represents prosumers' responsiveness to price variations. It shows self-elasticity for peak, off-peak, and valley periods on the main diagonal and cross-elasticity for non-diagonal arrays.

$$PEM = \begin{bmatrix} -0.1 & 0.016 & 0.012 \\ 0.016 & -0.1 & 0.01 \\ 0.012 & 0.01 & -0.1 \end{bmatrix} \qquad (23)$$

## 3. THE OPTIMIZATION APPROACH

This section will first present a two-way BEV charging optimization framework. Next, we will introduce an MOO method to address the problem. Finally, we will outline the decision-making strategy factoring in three conflict objective functions.

### 3.1 Framework

The implementation procedure of the BEV charging/discharging framework considering the compensation mechanism is depicted in Fig. 1. The core components include inputs, a two-way optimization framework, an MOO process, solution selection, and the resulting outputs. The data fed into the framework includes network data, application scenarios, and tariff structures outlining peak, off-peak, and shoulder hours. Moreover, collecting necessary data for probabilistic parameters, such as PV system generation data and driving behavior parameters, is essential. This comprehensive data gathering enables effective BEV allocation in each network node and EVCSs. Next, the two-way charging coordination block takes into account three conflicting objective functions: BEV owners' costs of exchange energy, network power losses, and EVCS profit, all while adhering to practical constraints such as power system limitations and limited space for each EVCS, limited distances for reaching EVCSs, etc. Then, the two-way BEV charging framework, comprising the blocks mentioned earlier, defined in section 2, is considered an MOO problem. The evaluation process iterates until the stopping condition of the algorithm is met. Additionally, a backward/forward load flow algorithm is utilized to ascertain the voltages of nodes [53]. Finally, the outputs include trade-off solutions delineating conflicting objective functions, a sequence of day-ahead charging and discharging decisions for each EVCS and BEV, and the resultant impact on grid operation. To evaluate the effectiveness of the proposed method in terms of flattening the load curve, reducing costs and losses, and maximizing benefits, three indices are employed: load factor (LF), peak to valley (P2V), and peak compensation (PC) [54]. The subsequent sections will delve into each step of the proposed method, comprehensively examining its intricacies and functionalities.

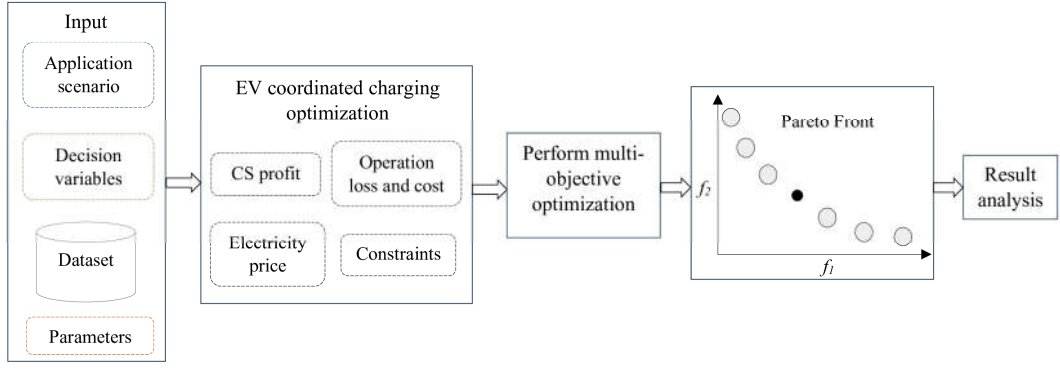

Fig. 1. The proposed multi-objective optimization schedule framework for EVCS

By applying Eq. (2) and the methodology in Fig. 2, we can allocate BEVs to each network node for every time slot. As the flowchart shows, we use the initial grid data to calculate the BEV owner's driving behavior, as defined in Eq. (1). $N^{CS}$ refers to the total number of EVCSs allocated in this study. BEVs are directed to the nearest charging station according to distance constraints, the capacity of the charging stations, and the price of electricity in the market. The chosen vehicle is only allocated to the charging stations if it satisfies the conditions. Otherwise, the BEV cannot participate in the program. Finally, each BEV is optimally scheduled to the EVCS for 24 hours.

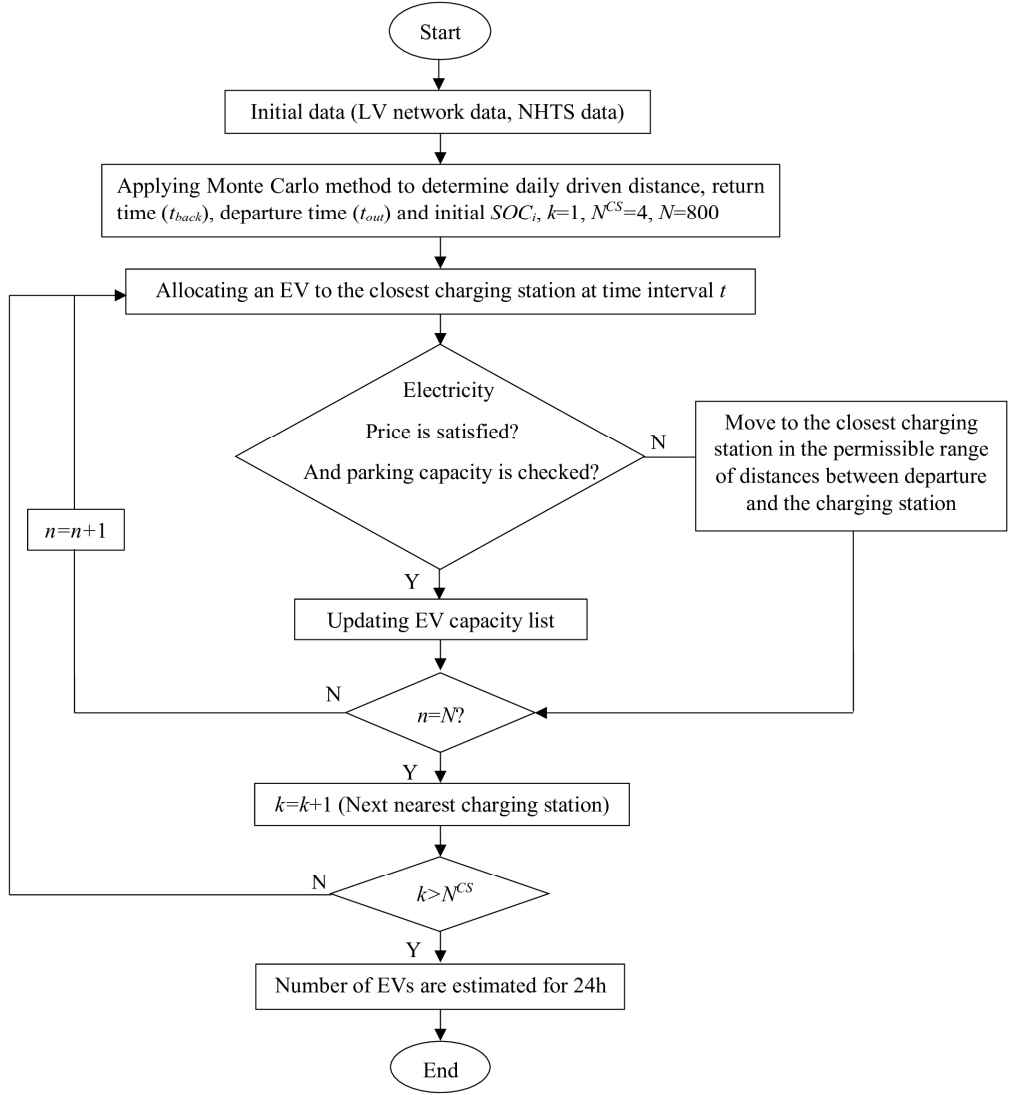

Fig. 2. The flowchart of the proposed BEV allocation method

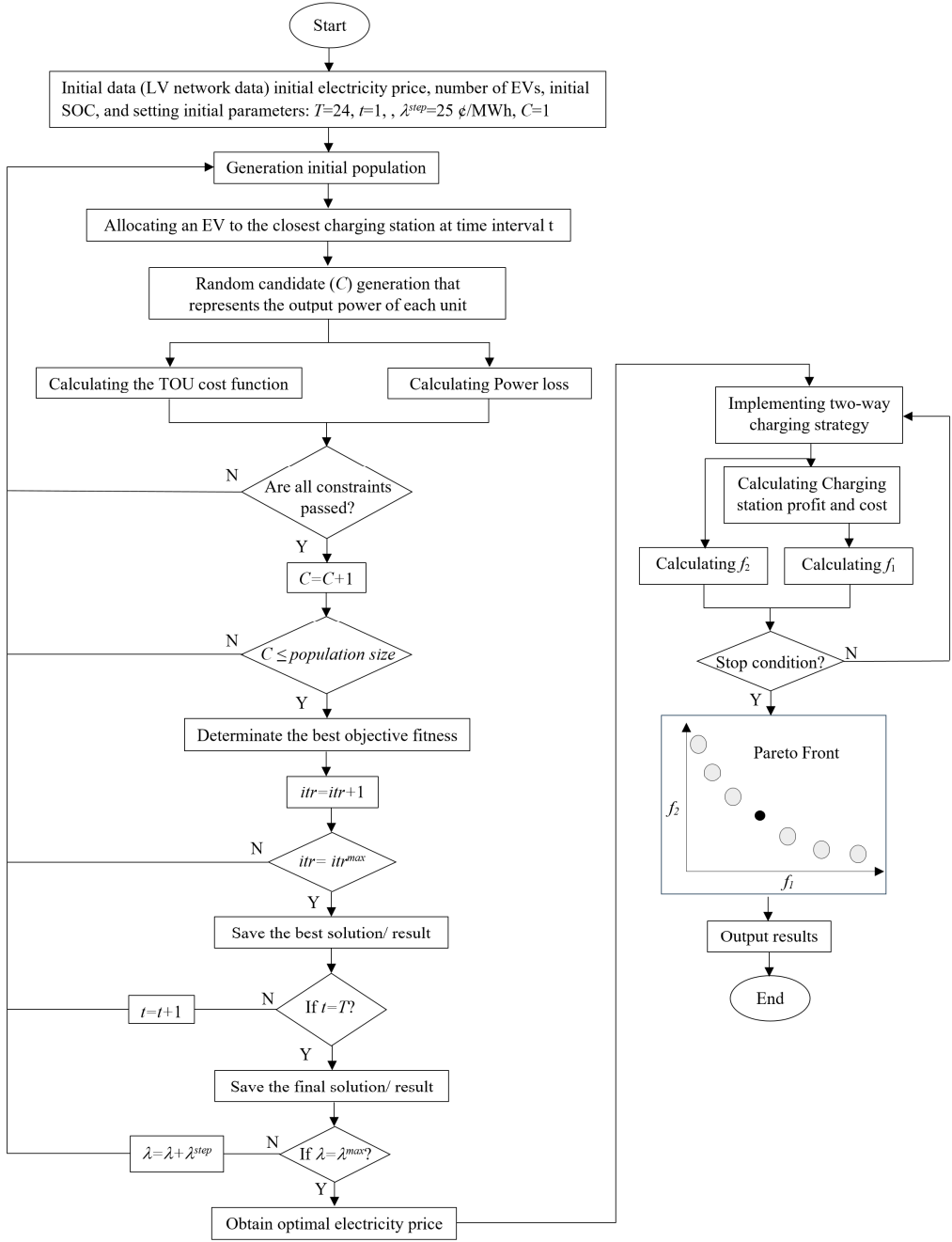

Fig. 3. The optimization process of the coordinated two-way charging strategy for EVCSs

### 3.2 Coordinated multi-objective BEV charging scheduling

Since BEV charging management, as mentioned in Eq. (3), is considered a linear problem, and DED is regarded as a non-linear problem, allowing for applying a suitable optimization method. To achieve the optimal charging/discharging profiles over 24 hours, we use the Non-dominated Sorting Genetic Algorithm II (NSGA-II) [47]. This algorithm is chosen for its high performance and ability to effectively address linear and non-linear optimization problems [71]. Based on our study, we compared NSGA-II, SQP, and the water-filling algorithm to evaluate their effectiveness in optimization for EV charging/discharging scheduling: i) NSGA-II outperforms both SQP and the water-filling algorithm by offering a more robust exploration of the solution space and effectively handling the stochastic behavior of BEVs, leading to more accurate solutions. ii) NSGA-II showed superior convergence and solution quality, consistently producing high-quality solutions near the Pareto front and maintaining solution diversity, while SQP converged more slowly with less diverse solutions, and the water-filling algorithm delivered less optimal results due to its heuristic approach. iii) NSGA-II's adaptability and flexibility made it better suited for dynamic and complex problems with less need for parameter tuning, while the water-filling algorithm, relying on fixed intervals for peak shaving and valley filling, lacked the flexibility to account for the variability in BEV behavior. Fig. 3 depicts the optimization process for a two-way charging strategy for EVCSs. Once the initial network, BEV data, and preliminary parameters have been obtained, the subsequent stage involves generating an array of potential solutions that capture diverse charging/discharging schedules for BEVs. Following this, the allocation of BEVs within each EVCS during individual time intervals is determined, considering the spatial limitations among EVCS locations. Through the incorporation of the load balance equation Eq. (22) and the identification of the candidate gas unit, the computation of TOU cost Eq. (16) and power losses Eq. (14) aids in establishing the optimal value of $\rho$. Consequently, this process prompts adjusting the TOU electricity price ($\lambda$), adopted as the optimized electricity price. Lastly, the execution of a two-way charging strategy is undertaken to minimize the cost for BEV owners at each EVCS, followed by the computation of two objective functions: maximizing EVCS benefit and minimizing power losses, as formulated in Eq. (10) and Eq. (14) respectively.

As illustrated in Fig. 3, the proposed planning requires a trade-off between EVCS profit and network loss cost after determining the Pareto-soptimal front, representing a range of solutions balancing these two objectives. However, for simplicity in comparing results across different scenarios, we apply a weighted factor to select one solution based on stakeholders' prior preferences instead of testing each feasible solution. The chosen solution can be determined using the following weighted criteria:

$$\min \quad f = \alpha f_1 + (1-\alpha) f_2 \qquad 0 \leq \alpha \leq 1 \tag{24}$$

where the $f_1$ and $f_2$ represent the EVCS profit and network losses, respectively, $\alpha$ refers to a weight for the total objective function, incremented by 0.01 from its initial value of 0.5.

## 4. THE SIMULATION RESULTS AND ANALYSIS

First, we provide the network details and related input data. Next, we present the optimization results of the NSGA-II Pareto front, including EVCS profit and network losses for each scenario. Additionally, we present each node's load profile, load factors, and voltages for selected scenarios over 24 hours. Finally, we conduct a sensitivity analysis on the price elasticity matrix and carbon credit prices, examining the impact of changing the charging/discharging rate on total cost and revenue.

Table 2: The modified BEV characteristics and types

| No. | BEV type | ECPK (kWh/km) | Participation level (%) | Battery capacity (kWh) | Maximal charging/discharging power (kWh) |
|---|---|---|---|---|---|
| 1 | Compact Sedan | 0.1625 | 60 | 10-20 | 0.8 battery capacity |
| 2 | Mid-Size Sedan | 0.1875 | 12 | 20-30 | 0.7 battery capacity |
| 3 | Mid-Size SUV | 0.2375 | 13 | 30-40 | 0.6 battery capacity |
| 4 | Full-Size SUV | 0.2875 | 15 | 40-50 | 0.5 battery capacity |

### 4.1 Dataset and network details

The modified IEEE 33 bus network is considered to exemplify the effectiveness of the proposed method, illustrated in Fig. 4. BEV data, including details such as BEV type, energy consumption per kilometer (ECPK), participation level of each BEV type in the model, battery capacity, and charging/discharging rate are shown in Table 2 [55, 56]. There are 800 BEVs in the three groups (A, B, and C). BEV charging stations are located at nodes 10, 21, 25, and 28. Ten DG units are considered at specific nodes [57]. Segments 1, 2, 3, 4, and 5, as shown in Fig. 4, are assigned to consumption pattern groups C, A, C, B, and A, respectively. The primary aim is to analyze and leverage various BEV load profile patterns based on statistical analysis, as detailed in [58]. In this study, the percentage of daily trips made by BEVs for different purposes is 34.3%, 30.2%, and 35.5%, respectively. The electricity tariffs during the valley period (00:00 am to 7:59 am) are 0.06 $/kWh, the price for the peak period (10 am to 1:59 pm and 6 to 9:59 pm) is 0.14 $/kWh, and the off-peak price (8 am to 9:59 am, 2 pm to 5:59 pm and 10 pm to 11:59 pm) is 0.1 $/kWh [59]. The daily load profile is considered for a working day in the summer peak that follows IEEE-RTS [55, 60]. To make the load profile suitable for our scenarios, we scale the daily load profile based on the IEEE-RTS spring season by a factor of 20. This adjustment is made to represent higher load demands. The modified profile ensures that voltages across all busbars remain within standard ranges compared to the 33 network load data snapshot, as shown in Fig. 5.

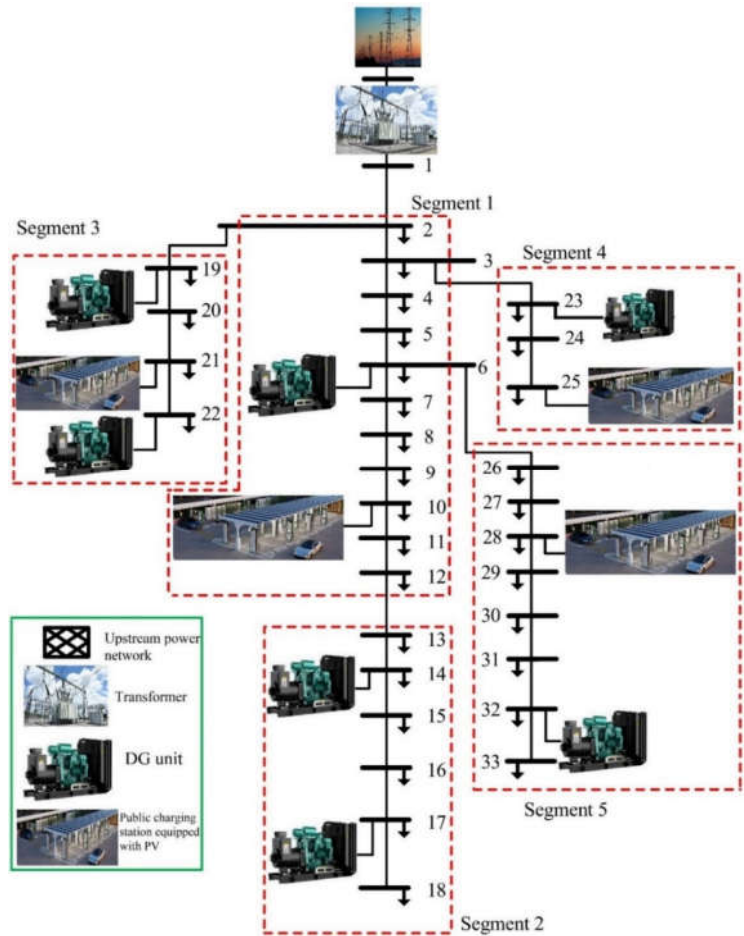

Fig. 4. The modified single-line diagram of the IEEE 33-bus test radial distribution grid [55]

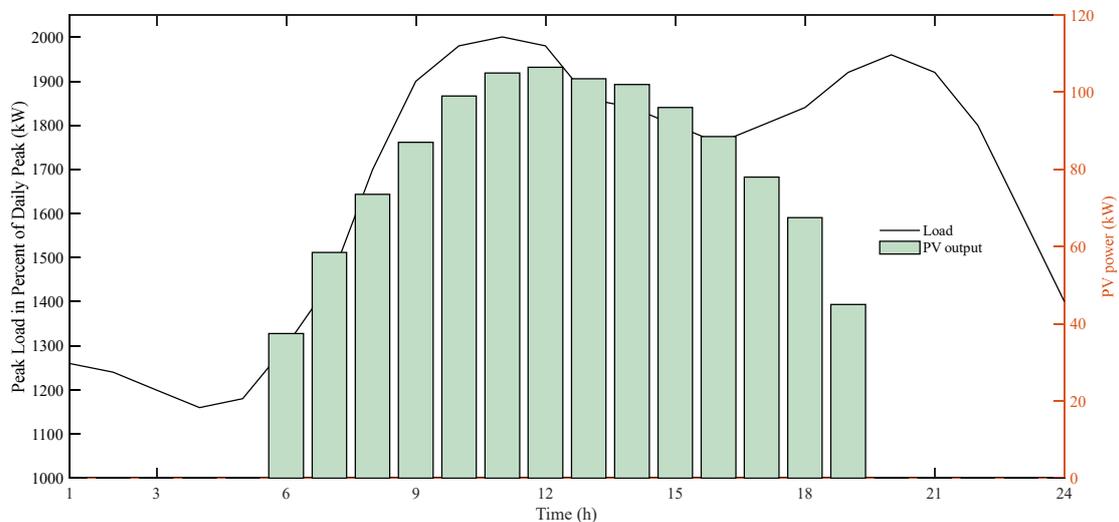

Fig. 5. IEEE-RTS hourly peak load in percent of a daily peak for the spring/fall season for working days [60] and PV generation

We utilize mono-crystalline modules for PV units in each charging [61]. The uncertainty of PV generation units is considered using the Monte Carlo method [62], shown in Fig. 5. The total number of BEVs within the network is 800, and the initial SOC of the BEVs is set as 15%-25%. The efficiency of charging/discharging operations is assumed to be 90%. The approach begins by introducing various scenarios, as outlined in Table 3, which are considered:

- S1 can be regarded as a baseline for comparison with scenarios that include management with V2G services.
- For scenarios S2-S7, some level of optimal management of coordinated charging/discharging is considered. It not only demonstrates the impact of these capabilities compared to the baseline scenario but also shows how they influence EVCS profits, BEV owner costs, network losses, voltage profiles, and load indices by comparing the results from different scenarios.

Table 3: Introduced scenarios

| Scenarios | V2G | BDC | RES | OEP | ERQ |
|---|---|---|---|---|---|
| S1 | ✗ | ✗ | ✗ | ✗ | ✗ |
| S2 | ✓ | ✗ | ✗ | ✗ | ✗ |
| S3 | ✓ | ✓ | ✗ | ✗ | ✗ |
| S4 | ✓ | ✓ | ✓ | ✗ | ✗ |
| S5 | ✓ | ✓ | ✗ | ✓ | ✗ |
| S6 | ✓ | ✓ | ✓ | ✗ | ✓ |
| S7 | ✓ | ✓ | ✓ | ✓ | ✓ |

All defined cases and computational analyses are conducted using MATLAB software, which is equipped with an Intel Core i7 2.6 GHz processor and 8 GB of RAM.

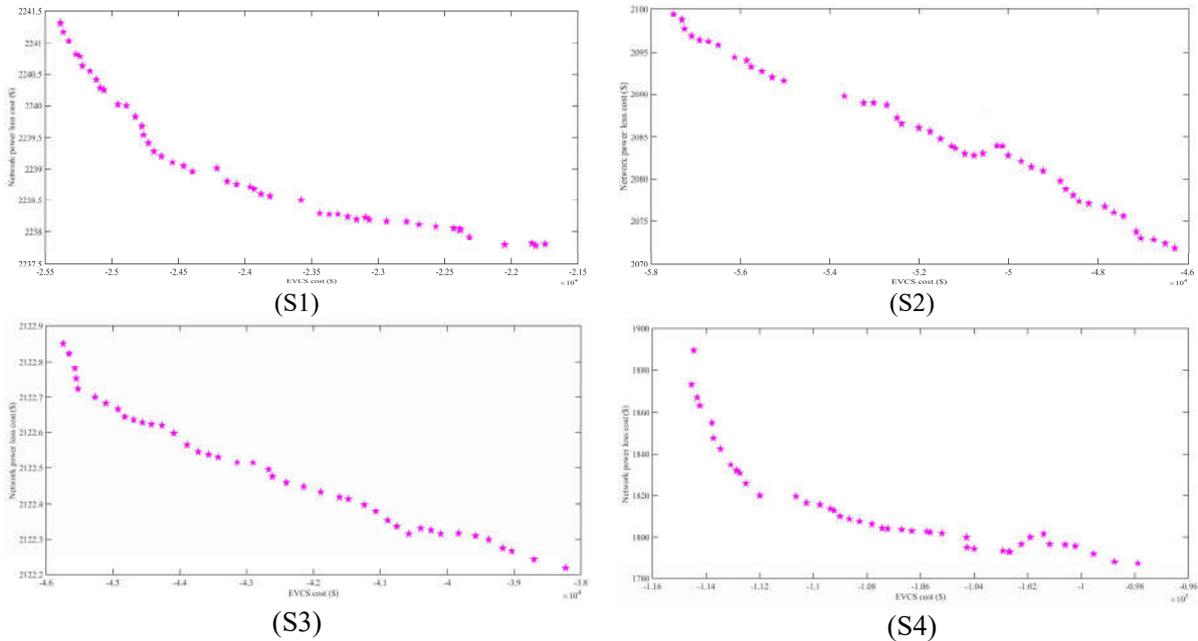

Fig. 6. Pareto fronts from NSGA-II on the total cost and CS profit of load over 24 hours for S1, S2, S3, and S4

### 4.2 Results

MOO aims to find a Pareto front with a set of trade-off feasible solutions for conflicting objectives. The results from the Pareto front present the range of objective functions, demonstrating how the proposed strategy balances different conflicting objectives, as shown in Fig. 6. Increasing EVCS benefits (or reducing costs) often leads to higher network power loss, negatively impacting grid voltage levels. Implementing an average TOU tariff allows for varying electricity prices based on charging/discharging requests and the base load. Optimal charging occurs during off-peak periods with lower tariffs, while optimal discharging occurs during peak periods with higher tariffs due to the TOU program changes. However, this strategy may decrease EVCS profit and reduce load variance, leading to a flattened load profile over time.

Table 4 presents the ranges obtained from the Pareto front using the NSGA-II algorithm. The Pareto front results are achieved by minimizing network losses and maximizing EVCS profit. Additionally, optimal electricity pricing mechanisms implemented in S5 help reduce the BEV user's cost and positively impact the load profile index, shown in Fig. 7.

Table 4: The objective function values on the Pareto front in ranges

| Scenario | This research | | SQP algorithm [25] | |
|---|---|---|---|---|
| | EVCS benefit ($f_1$) ×10$^3$ | Power grid losses ($f_2$) | EVCS benefit ($f_1$) ×10$^3$ | Power grid losses ($f_2$) |
| S1 | [2.15, 2.55] | [2237, 2241] | [1.9, 2.42] | [2250, 2260] |
| S2 | [4.6, 5.8] | [2070, 2100] | [4.4, 5.6] | [2080, 2100] |
| S3 | [3.8, 4.6] | [2112, 2192] | [3.5, 4.3] | [2130, 2205] |
| S4 | [9.6, 11.6] | [1780, 1900] | [8.5, 10.2] | [1800, 1920] |
| S5 | [3.74, 4.7] | [2172, 2218] | [3.5, 4.4] | [2180, 2220] |
| S6 | [4.46, 5.6] | [2281, 2338] | [4.2, 5.2] | [2190, 2405] |
| S7 | [5.51, 6.98] | [2407, 2483] | [5.01, 6.52] | [2400, 2490] |

As shown in Table 4 in the S1, the highest network losses are observed compared to other scenarios, as electricity demand spikes are met by supplying from the upstream network, leading to increased power losses. In contrast, for S2, introducing V2G services significantly increases the profit share to EVCS, approximately doubling it compared to S1. Additionally, the benefits to EVCS operators were evaluated based on the net profit derived from charging and discharging activities. This includes revenue from selling electricity back to the grid during peak hours, minus the operational costs, including energy procurement and infrastructure maintenance. We used a profit margin ratio to measure the increase in EVCS profitability. In addition to comparing our method with the SQP algorithm, we evaluated it against the MOO model presented in [25], which also focuses on stakeholder interests in coordinated two-way charging strategies. The comparison demonstrated that our proposed method offers stakeholders superior accuracy and more precise insights. This enhanced accuracy underscores the effectiveness of our approach in addressing the multi-objective optimization problem, ensuring better alignment with practical requirements and stakeholder needs. However, this approach negatively impacts the load profile and increases load variances, as shown in Fig. 7. Also, comparing S3 with S4, it is evident that by reducing the energy demand from the grid through the use of energy generated from PV systems installed in the EVCS, a higher benefit can be provided to EVCS owners, while also resulting in less network loss. This approach generates more income and benefits for EVCS owners while reducing power grid losses. By incorporating emission reduction programs in S6, we can expect several benefits for EVCS.

- Firstly, the program is expected to attract more BEV owners, improving power exchange trading between BEVs and the grid. This increased participation allows for more efficient utilization of EV batteries and optimized charging and discharging schedules.
- Secondly, the emission reduction programs provide additional financial incentives and sources of income for EVCS operators, which contribute to the overall revenue of the EVCS. As a result of these improvements, scenario S6 demonstrates a significant increase in benefits compared to S5. Specifically, there is an estimated 19% increase in benefits and 5% in power losses.
- Thirdly, the results indicate that introducing emission reduction programs not only contributes to environmental goals by reducing emissions but also enhances the economic viability and efficiency of EVCS operations. By balancing environmental sustainability and financial benefits, scenario S6 showcases a more comprehensive and desirable approach to BEV charging, positioning it as a win-win solution for stakeholders and the environment.

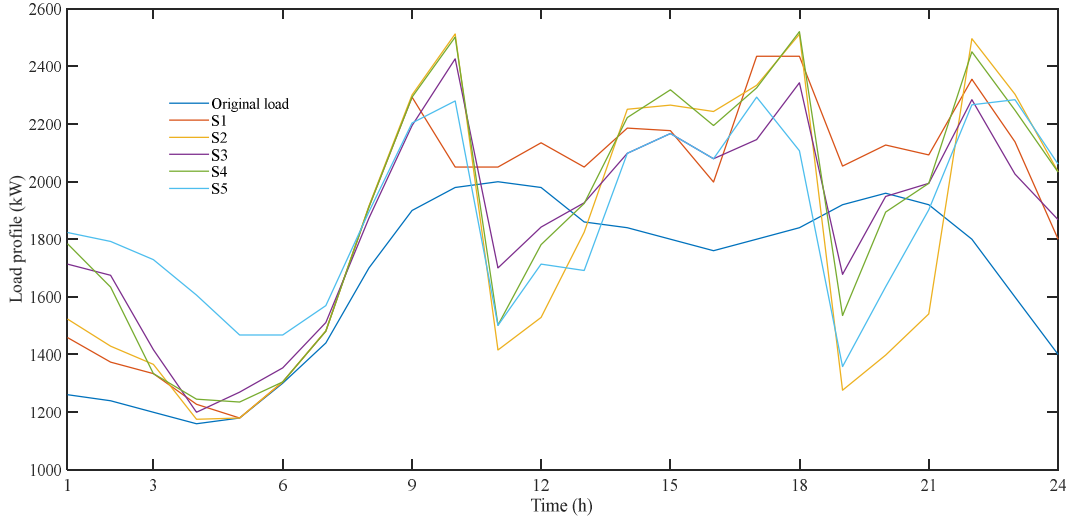

Fig. 7. Load profiling over baseload and other scenarios

Scenario S5 is to influence and optimize BEV users' consumption patterns by encouraging them to charge or discharge their BEVs more efficiently and cost-effectively. This is achieved through the implementation of an optimal pricing mechanism. The aim is twofold:

- Firstly, by obtaining the optimal electricity price, the study seeks to increase participation in the V2G service. This, in turn, helps reduce the load profile variance by incentivizing BEV owners to shift their charging time from peak to off-peak or valley periods. By aligning charging times with lower electricity tariffs, the overall electricity consumption can be better managed, leading to a more balanced grid load.
- Secondly, since BEV owners have diverse travel patterns and charging preferences, the study aims to find the optimal $\rho$-parameter that effectively motivates only about 20% of BEV users to join the V2G service, which is assumed in this study. This tailored approach considers individual driving characteristics and charging behavior, optimizing the incentive to suit the needs of a specific subset of BEV owners.
- Then, the results of S5 are illustrated in Fig. 7 and summarized in Table 5, the positive impact of the optimal pricing mechanism on BEV users' participation in the V2G service and its effects on load profile variance.

Table 5: Comparison of two scenarios, S3 and S5, applying different electricity pricing mechanisms

| Parameters | Value | |
|---|---|---|
| | S3 | S5 |
| Optimal $\rho$ ($/kWh) | - | 6.75 |
| Total power saved (kWh) | - | 287.281 |
| $C_{TOU}$ ($) | - | 98.851 |
| Total generation cost ($) | 1207.4912 | 910.6178 |
| Total cost ($) | 1207.4912 | 1009.4688 |

Table 6: Summary of load profile factors over scenarios

| Scenario | This research | | | Water-filling algorithm [41, 42] | | |
|---|---|---|---|---|---|---|
| | LF % | P2V% | PC% | LF% | P2V% | PC% |
| S3 | 76.82 | 50.1 | 0 | 72.5 | 55.08 | 0 |
| S4 | 81.74 | 40.78 | 5.81 | 75.1 | 52.96 | 6.26 |
| S5 | 78.64 | 47.69 | 2.39 | 73.9 | 49.11 | 4.13 |
| S6 | 79.97 | 45.36 | 4.07 | 76.3 | 46.68 | 7.44 |
| S7 | 86.97 | 36.14 | 6.28 | 83.4 | 41.24 | 8.35 |

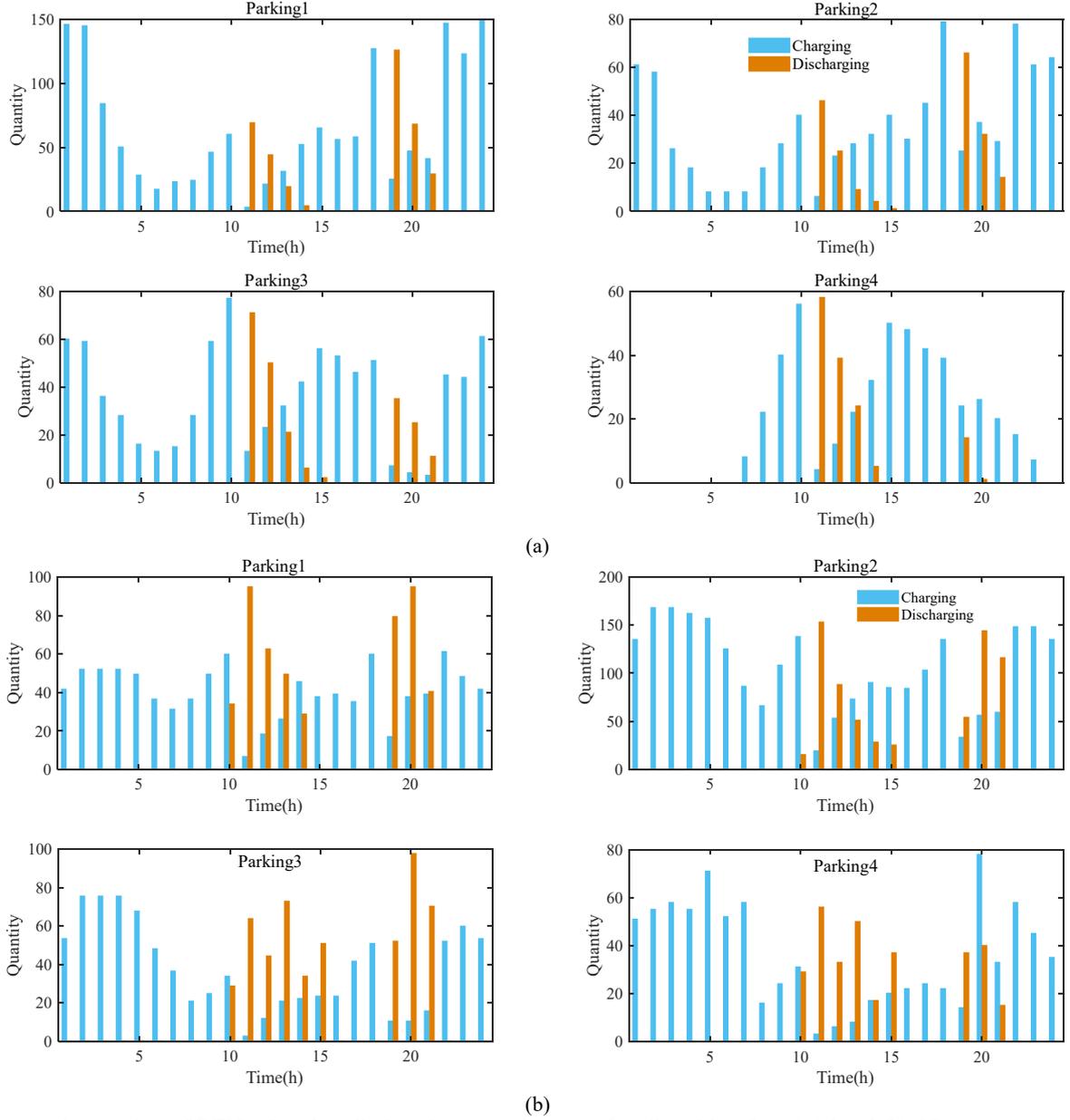

Fig. 8. The quantity number of BEVs charging/discharging over slow rate for all parking for (a) S6 and (b) S7

Table 5 compares two scenarios, S3 and S5, that involve different electricity pricing mechanisms. By optimizing the value for $\rho$ to be 6.75 $/kWh, a significant power saving of 287.281 kW is achieved over 24 h while considering the constraints of the EVCS, including its available capacity. We also examined how these strategies improved the overall load profile, reducing peak demand and smoothing the load curve. This was quantified using load factor metrics and peak-to-valley and compensation, indicating how effectively the load is distributed across time. Three indexes are provided in Table 6, indicating the impact of the strategy in minimizing load variance and flattening the load profile, as shown in Fig. 7. The load factor has improved from 76.82% in scenario S3 to 81.74% in S5, demonstrating the effectiveness of optimally scheduling the charging and discharging of BEVS based on optimal TOU tariffs. This leads to reduced network impact and improved peak-to-valley factor, which decreases by over 41%. The strategy encourages more prosumers to charge in off-peak times, helping to balance the network load and reduce the likelihood of overloading during peak hours. Besides, as seen in Table 6, the water-filling algorithm concentrates BEV charging during valley periods [1–4 am] and discharging during peak intervals [7, 8, and 10 pm]. In contrast, our approach's charging and discharging behaviors are more dispersed across various periods. This allows for more flexible use of charging infrastructure

in residential areas, providing BEV owners with greater convenience in charging/discharging during other intervals. Our method's optimal electricity price signal offers users more options, making the charging process more responsive to price variations. As a result, despite achieving similar grid load optimization, our dispatching process is more straightforward and practical, enhancing the overall significance of the results. Fig. 8 displays the charging and discharging quantities for all EVCSs at each time interval, focusing on a slow rate. Fig. 8 (a) highlights interesting patterns observed in three EVCS with more groups of A and B consumers. A prominent trend observed is the shift in charging times towards periods with more affordable electricity tariffs, characterized by lower electricity prices. This suggests BEV owners in these EVCS locations strategically leverage off-peak times to optimize their charging schedules.

Additionally, Fig. 8 shows that a significant portion of participation in the V2G service occurs during peak times. This suggests that BEV owners are actively utilizing battery capacity to discharge power back to the network when electricity demand is significant, helping to alleviate strain on the grid during peak hours. Furthermore, the demand for electricity in the late afternoon remains relatively high, presenting more opportunities for V2G services. This could be attributed to reduced generation from PV systems during sunset and consumers' continued need for electricity, making it advantageous for BEV users to join the V2G service during this time. By integrating an optimal electricity pricing mechanism with a carbon credit mechanism, V2G services can experience significant benefits, including financial incentives for BEV owners, reduced carbon emissions, and enhanced sustainability. S7 exemplifies the impact of this integrated approach on the load profile and its benefits for all stakeholders. As depicted in Fig. 8 (b), the number of BEVs participating in the V2G service increases compared to S6 due to the more favorable electricity prices. The optimal pricing mechanism leads BEV owners to adjust their charging schedules, moving charging activities from peak to valley times, thus optimizing grid load and reducing peak demand strain. We assessed the increase in V2G participation by comparing the number of BEVs engaged in V2G services across different scenarios. This was quantified by tracking the percentage of BEV users participating in V2G relative to the total number of BEV users.

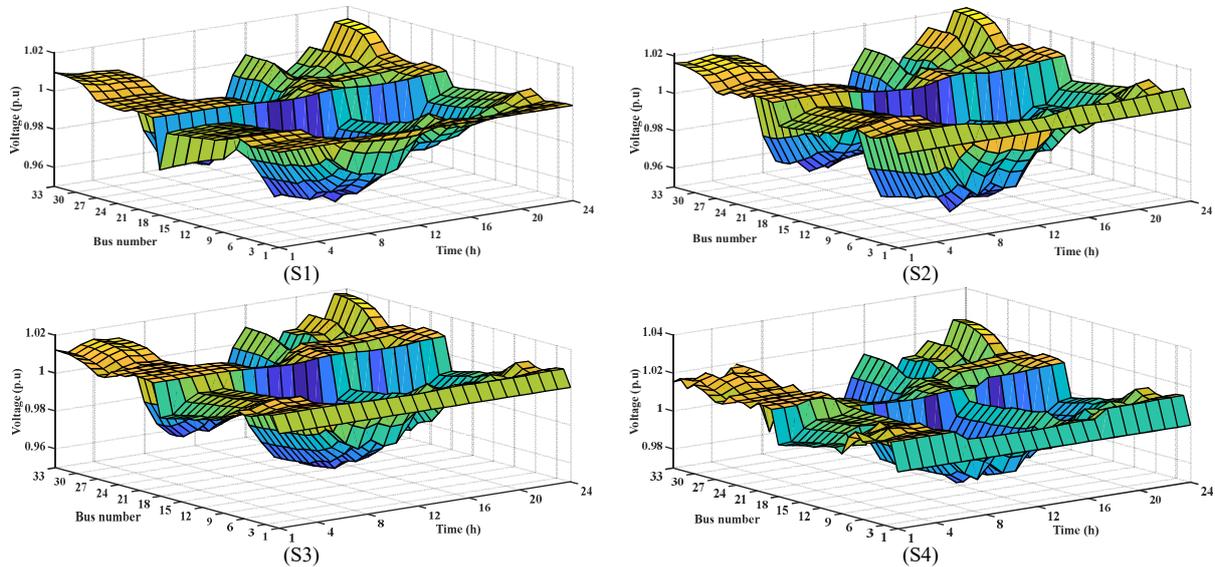

Fig. 9. Voltage profile of each node in different scenarios (S1-S4) over 24 hours

### 4.3 Voltage stability analysis

We also consider voltage profiles, aiming to meet power flow constraints. By installing ten DG units at specific busbars, the voltage profiles of all busbars, especially node 18 in the modified IEEE 33 bus system, significantly improve and remain above 0.95 per unit (p.u.). Fig. 9 depicts the voltage profiles of each node during various time intervals for different scenarios. Uncoordinated charging in scenario S1 results in node number 18 having the lowest voltage value of 0.95008 p.u. at 10 am and 0.96449 p.u. at 6 pm. With V2G capabilities in S2, the values increase to 0.95822 p.u. and 0.96832 p.u., respectively. In S3, node 18 has values of 0.96107 p.u. at 10 am and 0.96840 p.u. at 6 pm due to a drop in participants at certain times. However, in S4, with the injection of extra PV power into the grid and reduced power flow from PV generation, the grid exhibits a more stable voltage profile. For example, node number 18 reaches 0.98839 p.u. and 0.997308 p.u. at 10 am and 3 pm, respectively. This demonstrates the strategy's effectiveness in improving voltage stability and meeting power system criteria.

### 4.4 Sensitive analysis

Critical variables are considered to investigate the proposed framework's performance. We aim to explore how the introduced strategies respond to changes in some factors and assess their impact on the overall results. By exploring different scenarios and adjusting key parameters, we can gain insights into the approach's robustness and effectiveness.

#### 4.4.1 Price elasticity matrix analysis

The sensitivity analysis for two additional groups with 0.5 and 2 times the PEM compared to the baseline scenario S5. S5 is discussed, which will demonstrate how variations in PEM and driving patterns can significantly impact the performance of the proposed strategy. We refer to these groups as S5 (a) and S5 (b), respectively, for comparison purposes. In scenario S5 (a), following the implementation of the TOU program, electricity prices surge during the daily peak period, driven by the charging behaviors of BEV owners and their initial SOC. Consequently, EVCS benefits are higher when utilizing the V2G capability, as shown in Table 7. During the valley period, when electricity prices are at their minimum value, the EVCS benefit decreases. Thus, more income is generated during the day than at night, resulting in higher overall profit.

Table 7: Sensitivity analysis of implementing TOU in different PEM

| Parameters | Value (first BEV charging strategies) | | |
|---|---|---|---|
| | S5 | S5 (a) | S5 (b) |
| Optimal $\rho$ ($/ kWh) | 6.75 | 17.25 | 3 |
| Total power saved (kWh) | 287.281 | 170.189 | 395.67 |
| $C_{TOU}$ ($) | 98.851 | 61.98 | 174.88 |
| Total generation cost ($) | 910.6178 | 813.21 | 953.067 |
| Total cost | 1009.4688 | 875.19 | 1127.947 |
| Load Factor (LF) % | 81.74 | 83.34 | 81.64 |
| Peak to Valley (PV) % | 40.78 | 38.26 | 40.88 |
| Peak Compensation (PC) % | 5.81 | 6.56 | 5.11 |
| EVCS benefit ($f_1$) ×$10^3$ | [3.74 – 4.7] | [3.98 – 5.0] | [3.16 – 4.08] |
| Power loss ($f_2$) | [2172 – 2218] | [2101 – 2175] | [2172 – 2218] |

S5(b) reverses the situation, decreasing total income. The TOU program causes the electricity price to be lower during the daily peak period for this group, affecting EVCS benefits in the opposite direction. Regarding grid losses, S5(a) shows reduced total losses due to the implementation of the TOU program and power-saving measures. On the other hand, in S5(b), the opposite occurs, and the losses increase. By understanding these sensitivities, decision-makers can tailor the implementation of optimal pricing mechanisms and V2G services better to suit specific BEV owner groups and grid conditions, optimizing the overall benefits and sustainability of the system.

#### 4.4.2 Carbon credit mechanism prices analysis

Carbon trade price is crucial for the program's success. A higher price generates more revenue to offset EV battery costs. However, sufficient compensation to incentivize BEV adoption must be balanced against overcompensation, which could increase system costs. The results can help identify the best carbon price that ensures adequate compensation for battery degradation while promoting BEV adoption and V2G participation. Hence, we set the carbon trade price by step +0.03 $/kg and conduct the analysis based on scenario S7, which offers remarkable insights into how carbon trading prices could offset the costs of EV battery degradation. Fig. 10 shows that higher carbon prices generally lead to increased revenues from carbon trading. This revenue can be a valuable funding source to compensate for EV battery degradation costs, providing a strong incentive for consumers to adopt BEVs. By offering a means to recover some of the battery replacement expenses over time, the compensation program becomes an attractive proposition for potential BEV owners. In this study, we suppose EVCS owners allocate their carbon revenue as incentives for BEV owners. The effectiveness of the compensation program becomes even more promising. With a carbon credit price of $0.15, it is expected that using a carbon credit mechanism could almost fully compensate for battery degradation costs. This significant incentive has the potential to accelerate the adoption of BEVs, especially in the context of V2G services, where BEV users can participate in carbon trading and contribute to reducing carbon emissions. Combining the carbon trading benefits and EV battery degradation compensation creates a compelling proposition for consumers to embrace BEVs.

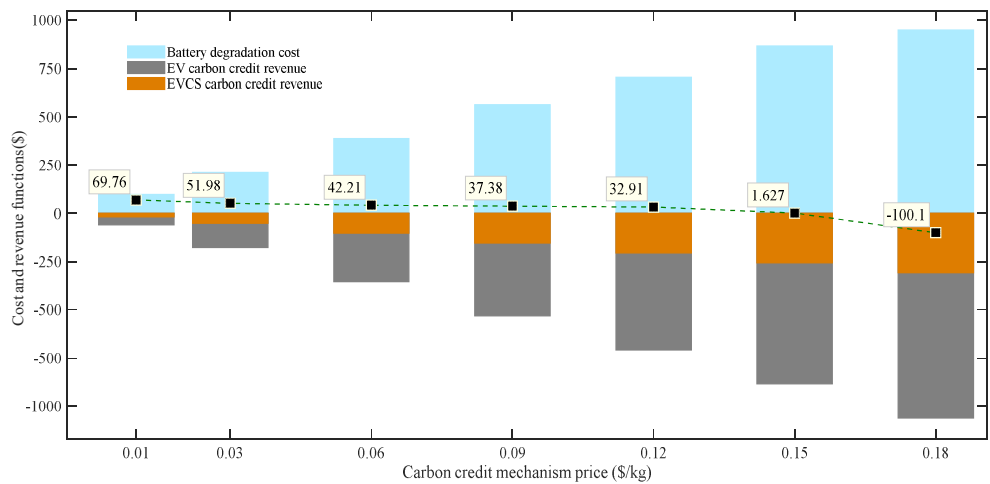

Fig. 10. Sensitivity analysis on carbon credit prices

### 4.4.3 Charging/discharging rate analysis

An EV battery's charging and discharging rate can significantly influence various aspects. These include the expenses borne by the BEV owner due to battery degradation costs, the advantages of an EVCS, and the mitigated power losses. The rapid charging and discharging of BEVs can impact the power grid, affecting peak demand, voltage oscillations, grid stability, and energy losses. So, in this section, we assign three distinct charging/discharging rates: slow, regular, and fast to assess the operational effectiveness of the proposed method and strategy in achieving optimal outcomes. We also analyze the determined functions to understand their trend over time.

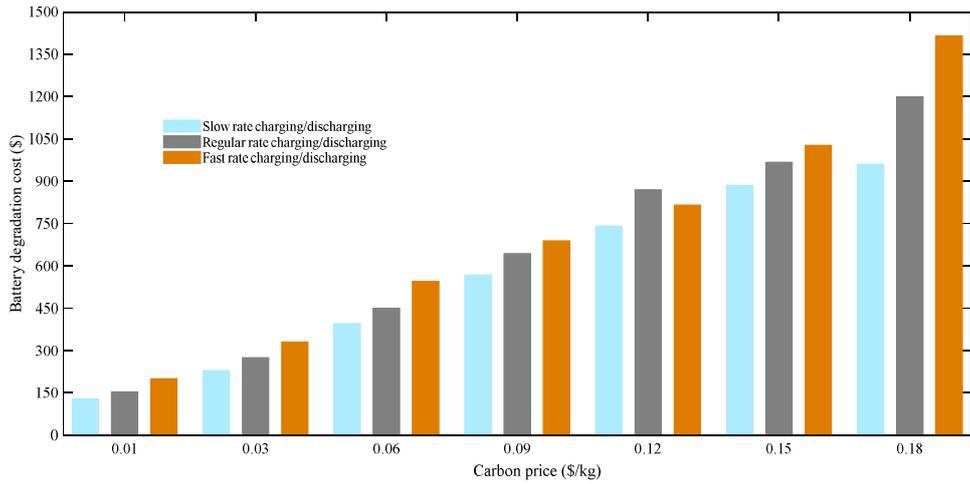

Fig. 11. Sensitivity analysis on battery degradation cost vs. carbon credit prices under different charging/discharging rates

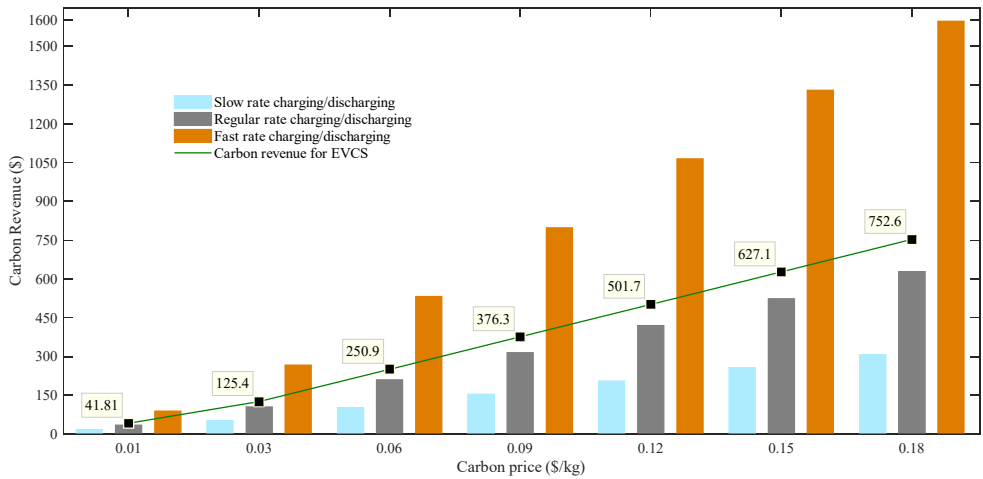

Fig. 12. Sensitivity analysis on carbon revenue vs. carbon credit prices under different charging/discharging rates

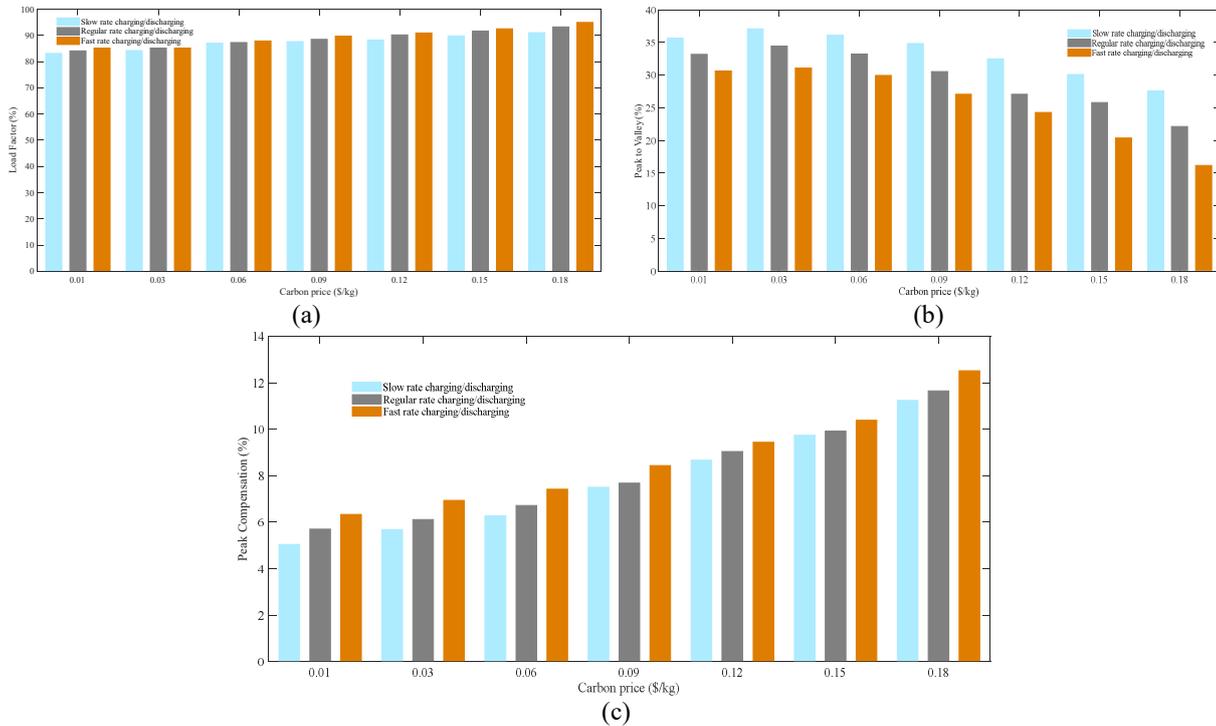

Fig. 13. Sensitivity analysis on carbon revenue vs. (a): load factor, (b): peak valley, (c): peak compensation under different charging/discharging rates

Fig. 13. illustrates how alterations in the BEV charging/discharging rate within a coordinated two-way charging initiative can influence the load profile index. Notably, a swifter charging/discharging rate translates to reduced charging time for the BEV, offering convenience to the owner and creating more opportunities for additional BEVs to charge due to the limited capacity of EVCSs. Consequently, a more significant number of BEV owners can participate in the program, leading to a positive effect on the grid's load profile index, as depicted in Fig. 13. For example, the transition from a regular to a fast charging/discharging level, at a specific carbon price of 9 cents, yields a favorable impact on the load profile (Fig. 13(a)). The load profile improves from 82% during regular charging to 84.5% during fast charging.

## 5. CONCLUSIONS, LIMITATIONS, AND FUTURE WORKS

### 5.1 Conclusions

In this paper, we developed a multi-objective coordinated BEV charging optimization framework that incorporated two cost reduction programs: a carbon emission reduction program and dynamic economic dispatch combined with a demand response program. The framework simultaneously addressed three competing objectives: maximizing EVCS profit, minimizing network power losses, reducing BEV owners' costs, and achieving a balanced optimization that catered to all stakeholders' needs. We integrated diverse BEV load profiles by considering different driving patterns and varying charging/discharging rates. This ensured that the model reflected real-world scenarios while capturing uncertainties in load profiles and PV generation. Additionally, the study implemented dynamic economic dispatch and TOU strategies to generate optimal electricity pricing signals. These strategies encouraged BEV owners to participate in V2GG services, optimizing their load profile patterns and resulting in cost reductions and maximized benefits. Lastly, we introduced a novel compensation mechanism for carbon credits to address battery aging costs, incentivizing both BEV and EVCS owners to participate in V2G services and aligning economic benefits with environmental sustainability. Through seven distinct scenarios implemented on a modified IEEE 33-bus system with 800 BEV users, our results demonstrated several key findings:

- **Increased engagement in G2V and V2G services**: By increasing the carbon credit price by 14%, BEV owners' concerns about battery degradation costs are nearly eliminated, encouraging more participation in V2G services. This results in improved load profiles and full coverage of battery degradation costs, making V2G services more attractive and beneficial for BEV owners and the grid.

- **Economic benefits**: By setting the weight value of objective functions at 50% and implementing full control over BEV loads, including cost reduction programs and V2G services, the proposed scheduling method nearly doubles the EVCS benefits compared to uncontrolled management. Additionally, it reduces power losses by 8.5%, demonstrating the coordinated BEV charging strategy's significant positive impact on economic and operational performance.

- **Compensation for battery degradation**: By increasing the charging/discharging rate from slow to fast at a specific carbon price, such as 0.12 S/kg, battery degradation costs rise by 2.7%. However, the revenue generated from the carbon credit mechanism significantly increases from $165 to $1,047, driven by a higher rate of participation in V2G services. This shift highlights the financial benefits of faster charging rates despite the associated increase in battery costs.

- **Impact of price elasticity**: Comparing different elasticity and consumption profiles, reducing the elasticity matrix by 50% resulted in a 17.25 $/kWh improvement in the value of $\rho$. This led to significant enhancements in load management, with the load factor improving by 1.96% and the peak-to-valley load compensation index increasing by 6.2% and 12.9%, respectively. These changes contributed to more efficient energy distribution and better load balancing.

### 5.2 Limitations

This study utilized a test network to evaluate the proposed methods, offering valuable insights while acknowledging limitations. The test network is a simplified version of real-world power systems, which may not fully capture the complexities of actual networks, including intricate topologies and component interactions. The load profile is also based on a specific spring season scenario, which may not represent varying conditions in other regions or seasons. Real-world networks experience dynamic operational changes that may affect the system's performance in ways not reflected in the study. Moreover, the results may not scale directly to larger, more complex networks, and real-world factors like economic and regulatory constraints could impact the applicability of the findings. Further investigation is required to assess scalability and real-world relevance.

### 5.3 Future work

Future research should focus on several key areas to address these limitations and enhance the applicability of our findings. First, validating the proposed methods across various real-world networks is crucial. This involves testing the methods on networks with different topologies, load profiles, and operational conditions to assess their robustness and adaptability to diverse scenarios. Expansion planning is another critical area for future research. As BEV adoption increases, effective expansion planning will be necessary to ensure the reliability of the power network. This requires collaboration among utilities, regulators, and other stakeholders to address challenges and maintain grid stability. Furthermore, given the lack of comprehensive data and limited public charging stations in Australia, gathering actual network data for testing and refining these approaches is essential. This will help to better align the proposed solutions with real-world conditions. Exploring the integration of peer-to-peer (P2P) scenarios is also recommended. While the current study focuses on EVCSs, examining home charging facilities and P2P interactions could provide additional insights.

### CRediT authorship contribution statement

**Saman Mehrnia:** review & editing, writing – original draft, Software, Visualization, Validation, Methodology, Investigation, Formal analysis, Data curation, Conceptualization. **Hui Song:** review & editing. **Nameer Al Khafaf:** review & editing. **Mahdi Jalili:** Writing – review & editing, Project administration, Investigation, Funding acquisition. **Lasantha Meegahapola:** Project administration, Investigation, Funding acquisition. **Brendan McGrath:** Project administration, Investigation, Funding acquisition.

### Declaration of competing interest

The authors declare that they have no known competing financial interests or personal relationships that could have appeared to influence the work reported in this paper.